\documentclass[twocolumn,showpacs,preprintnumbers,
amsmath,amssymb]{revtex4}
\usepackage{graphicx}
\usepackage{multirow}

\def\beq{\begin{equation}}
\def\eeq{\end{equation}}
\def\bea{\begin{eqnarray}}
\def\eea{\end{eqnarray}}
\def\ba{\begin{array}}
\def\ea{\end{array}}
\def\bit{\begin{itemize}}
\def\eit{\end{itemize}}
\def\pa{\partial}
\def\na{\nabla}
\def\nn{\nonumber}

\def\al{\alpha}

\def\Ga{\Gamma}
\def\si{\sigma}

\def\de{\delta}

\def\la{\lambda}

\def\na{\nabla}
\def\rd{{\rm d}}

\def\lPL{\ell_{\rm Pl}}

\def\bx{{\bf{x}}}
\def\by{{\bf{y}}}

\def\Feff{{\Gamma}_{\rm anom}}
\def\rd{{\rm d}}

\def\cK{{\cal K}}

\def\trad{t_{\rm rad}}

\def\GB{{{\rm E}_4}}

\makeatletter
\def\underbracket{%
  \@ifnextchar [ %
    {\@underbracket}%
    {\@underbracket [\@bracketheight]}}
\def\@underbracket[#1]{%
  \@ifnextchar [ %
    {\@under@bracket[#1]}%
    {\@under@bracket[#1][0.4em]}}
\def\@under@bracket[#1][#2]#3{%\message {Underbracket: #1,#2,#3}
  \mathop {%
    \vtop {%
      \m@th \ialign {%
        ##\crcr $\hfil \displaystyle {#3}\hfil $%
       \crcr \noalign %
       {\kern 3\p@ \nointerlineskip }%
        \upbracketfill {#1}{#2}
       \crcr \noalign %
       {\kern 3\p@ }%
     }%
   }%
  }%
  \limits%
}
\def\upbracketfill#1#2{%
  $\m@th \setbox \z@ \hbox {$\braceld$}
  \edef\@bracketheight{\the\ht\z@}\bracketend{#1}{#2}
  \leaders \vrule \@height #1 \@depth \z@ \hfill
  \leaders \vrule \@height #1 \@depth \z@ \hfill%
  \bracketend{#1}{#2}$%
}
\def\bracketend#1#2{\vrule height #2 width #1\relax}
\begin{document}
\title{Conformal Anomalies and Gravitational Waves}
\author{Krzysztof A. Meissner$^1$ and Hermann Nicolai$^2$}
\affiliation{$^1$Faculty of Physics, University of Warsaw\\
ul. Pasteura 5, 02-093 Warsaw, Poland\\
$^2$Max-Planck-Institut f\"ur Gravitationsphysik
(Albert-Einstein-Institut)\\
M\"uhlenberg 1, D-14476 Potsdam, Germany\\
}

\begin{abstract} 
We argue that the presence of conformal anomalies in gravitational theories can lead 
to observable modifications to  Einstein's equations via the  induced anomalous 
effective actions, whose non-localities can overwhelm the smallness of 
the Planck scale. The fact that no such effects have been seen in recent 
cosmological or gravitational wave observations therefore imposes strong restrictions 
on the field content of possible extensions of Einstein's theory: all viable theories 
should have vanishing conformal anomalies. We then show that
a complete cancellation of conformal anomalies in $D=4$ for both the 
$C^2$ invariant and the Euler (Gauss-Bonnet) invariant $\GB$ can only
be achieved for  $N$-extended supergravities with $N \geqslant 5$, as well 
as for M theory compactified to four dimensions.  
\end{abstract}

\maketitle

\vspace{0.1cm}
\noindent{\bf 1. Introduction.} In this Letter, we study the effect of conformal anomalies
on low energy (sub-Planckian) gravitational physics. While the effect of 
the non-anomalous part of the non-local effective action on Einstein's equations for 
any matter coupling is completely negligible away from the Planck regime, due to 
the powers of $\lPL$  (the Planck length) that come with higher order curvature 
and other quantum corrections, we will here argue that the {\em anomalous part} $\Feff[g,\phi]$ 
of the non-local effective  action, which is responsible for the conformal anomaly via formula 
(\ref{Fanom}) below, is not only relevant for the quantised theory, but could also affect
classical low energy processes  (such as the propagation of gravitational waves) 
in observable ways, for instance by exciting longitudinal gravitational degrees of freedom.
This effect is made possible through the compensation of the smallness of the Planck scale 
$\lPL$ by a large factor coming from the non-locality, analogous to the standard
axial anomaly which arises by the competition of a UV divergence and a 
vanishing expression in $D=4$ ({\em e.g.} in dimensional regularisation).
It can thus be viewed as a low energy (IR)  manifestation of trans-Planckian physics. 
The absence of such modifications in recent cosmological or gravitational wave
observations leads us to conclude that all conformal anomalies must cancel -- in analogy
with the cancellation of chiral gauge anomalies in the Standard Model of particle physics,
required for the latter's consistency and renormalizability. Our considerations apply
to any context where Einstein's equations play a central role ({\it e.g.} black holes, dispersion relations for gravitational waves,  amplification of primordial quantum fluctuations, {\it etc.}).

\noindent{\bf 2. Conformal Anomaly.} We refer to \cite{Deser2,Duff1,CD,FT,DS,Deser1,T} 
for a derivation and basic properties 
of the conformal anomaly, as well as for further references. The conformal anomaly has
two sources, namely the fact that the UV regulator for any conformal matter system 
coupled to gravity necessarily breaks conformal invariance, and secondly the fact that 
even for classically conformal fields the functional measure depends on the metric in a 
non-local manner. In four dimensions  the anomaly takes the form 
\cite{Deser2,Duff1,CD} 
\beq
{\cal A}=T^\mu_{\; \; \mu}=\frac{1}{180(4\pi)^2}\left(c_{s}C^2+a_s \GB \right)
\label{anom}
\eeq\\[-4mm]
where 
\bea\label{CGB}
C^2&\equiv&C_{\mu\nu\rho\si}C^{\mu\nu\rho\si}=R_{\mu\nu\rho\si}
R^{\mu\nu\rho\si}-2R_{\mu\nu}R^{\mu\nu}+\frac13 R^2\nn\\
\GB &=& R_{\mu\nu\rho\si}
R^{\mu\nu\rho\si}-4R_{\mu\nu}R^{\mu\nu}+R^2
\eea
and the coefficients $c_s$ and $a_s$ depend on the spin $s$ of the field that
couples to gravity, and on whether these fields are massless or massive, see Table below.
$\GB$ is the Gauss-Bonnet density, a total derivative that gives a topological invariant
when integrated over a 4-dimensional manifold.

While the anomaly (\ref{anom}) itself is a local expression, 
it is well known that it cannot be obtained by variation of a local action functional 
(otherwise the anomaly could be cancelled by a local counterterm). 
Accordingly,  we are here concerned with the {\em anomalous} part $\Feff[g,\phi]$ of the 
non-local effective action that gives rise to (\ref{anom})  via variation of the 
conformal factor,
\beq\label{Fanom}
{\cal A} (x) = 
- \frac2{\sqrt{-g(x) }} g_{\mu\nu}(x) \frac{\delta \Feff [g,\phi]}{\delta g_{\mu\nu} (x)} 
\eeq
where $\phi$ stands for any kind of matter field.
In four dimensions such actions -- apart from not being unique -- are not known  in 
completely explicit form, although there are partial results \cite{DS,Deser1,Mott,Mott1}. This is
in contrast to the case $D=2$ where there {\em is} a unique expression satisfying all 
consistency requirements, namely the famous Polyakov formula 
$\Feff^{D=2}[g] =-\frac12 \int \sqrt{-g}R \, \Box^{-1}_g R$,
where $\Box_g^{-1}$ is the Green's function \cite{P}.
As is very well known \cite{P,DK},  the presence of this anomalous term in the 
effective action changes the behaviour of the quantised theory in dramatic 
ways --  altering the critical dimension of the effective string theory and adding a 
new propagating (Liouville)  degree of freedom to restore quantum conformal 
invariance. For higher dimensional gravity the {\em non-local} effects of $\Feff[g,\phi]$ 
have not been much discussed so far (but see \cite{Davies,Hartle,Mott}), not only 
because of the absence of sufficiently explicit expressions, 
but also for the obvious reason that gravity is non-renormalizable, whence the
further addition of anomalous vertices would only make bad things worse. 
By contrast, for a finite or renormalizable theory of quantum gravity the extra term  
does make a difference, like for Yang-Mills theories where anomalies are well 
known to spoil the renormalizability, hence the consistency, of the quantised theory.

\vspace{0.2cm}
\noindent{\bf{3. Classical considerations.}} 
We now show that the presence 
of an anomalous contribution $\Feff$ to the effective action {\em may affect the classical 
theory in observable ways}. The addition of $\Feff$ to the effective
action leads to an effective modification of Einstein's equations, {\it viz.}
\beq\label{MEFT}
\lPL^{-2} \left(R_{\mu\nu} -\frac12 g_{\mu\nu} R \right)(x) =  
\,- \, \frac2{\sqrt{-g}} \frac{\delta \Feff [g,\phi]}{\delta g^{\mu\nu}(x)}  + \cdots 
\eeq
where the dots stand for non-anomalous matter contributions as well as contributions 
from non-anomalous higher order curvature corrections (which are negligible).
The above equation entails the consistency condition 
\vspace{-1mm}
\beq\label{DivF}
\na^\mu \left(
\, \frac2{\sqrt{-g(x)}} \frac{\delta \Feff [g,\phi]}{\delta g^{\mu\nu}(x)} \right) \,=\, 0
\eeq
where the vanishing divergence of the contribution from the non-anomalous part
of the effective action follows from the standard Ward-Takahashi identities. 

Our main interest here is in the variation of $\Feff$ w.r.t. metric deformations which are  
{\em neither} trace {\em nor} diffeomorphisms; we may generically refer to such 
deformations as `gravitational waves' (in the broad sense; the proper gravitational waves must 
satisfy the conditions at infinity first identified by Trautman \cite{Tr}).  Unlike for (\ref{Fanom}), 
such variations cannot be expected to be local, as is already evident for the  Polyakov action
for which
\bea\label{Pol2}
\frac{2}{\sqrt{-g}}\frac{\de \Feff^{D=2}[g]}{\de g_{\mu\nu}} &=&
2g^{\mu\nu}R - 2\nabla^\mu\nabla^\nu\left(\frac{1}{\Box_g} R\right)   \\[1mm]
&& \!\!\!\!\!\!\!\!\!\!\!\!\!\!\!\!\!\!\!\!\!\!\!\!\!\!\!\!\!\!\!\!\!\!\!\!\!\!\!\!\!\!\!
+ \, (g^{\mu\al}g^{\nu\beta}-\frac12 g^{\mu\nu}
g^{\al\beta})\na_\al \left(\frac{1}{\Box_g} R\right)
 \na_\beta \left(\frac{1}{\Box_g} R\right)          \nn
 \eea
Tracing with $g_{\mu\nu}$ we recover the standard  conformal anomaly; also, this expression 
does satisfy (\ref{DivF}) identically. This result illustrates our claim that  variations 
of $\Feff$ other than trace variations remain  non-local. Equally important, it exhibits potential 
singularities associated with the Green's functions $\Box_g^{-1}(x,y)$.  However,  
in this particular case the singularities are harmless because of a well known 
peculiarity of $D=2$ gravity:  there are no deformations of the metric other than 
conformal or diffeomorphism variations \footnote{In Liouville theory 
with {\em Euclidean} signature with compact Riemann surfaces as worldsheets all singularities are actually integrable because the scalar propagators only exhibit logarithmic singularities 
at coincident points}.

For $D=4$, matters are much more involved, because the metric now carries
propagating degrees of freedom. In this case $\Feff$ will  be a very complicated 
functional, in general involving an infinite series of products 
of Riemann tensors interspersed with massless propagators in some gravitational 
background.  Even if (\ref{DivF}) is satisfied, hence diffeomorphism invariance 
maintained, the example of Liouville theory teaches us that $\Feff$ must be 
expected to excite {\em new} (longitudinal) gravitational degrees of freedom. 
In quantum gauge theory, this effect would entail the breaking of gauge invariance 
via the insertion of anomalous triangles into vector boson self-energy diagrams. 
In addition, the presence of $\Feff$ can lead to observable consequences even in 
the classical approximation, provided we can show that the smallness of $\lPL$ 
can be overcome by the non-localities on the r.h.s. of (\ref{MEFT}).

To make all this a little more explicit we observe that, in lowest order, $\Feff^{D=4}$ must 
contain a term $\propto\cK \,\Box_g^{-1}R$ (also present in the anomalous effective action for 
$\GB$ proposed in \cite{DS}), where $\cK\equiv R_{\rho\si\la\tau}R^{\rho\si\la\tau}$ is the 
Kretschmann scalar \footnote{Similar non-localities are present for the other terms
exhibited in \cite{DS}, as well as for the nonlocal actions involving inverse quartic differential 
operators proposed in \cite{Deser1,Mott}.}. In analogy with (\ref{Pol2}), the Einstein equation  
(\ref{MEFT})  will thus be modified by a term proportional to 
\beq\label{DesSch}
\propto \, 
 \na_\mu \left( G \star\cK \right)   \na_\nu \left( G \star R \right)
\; + \; (\mu\leftrightarrow \nu)
\eeq
where $G$ is the Green's function associated with the conformal d'Alembertian,  {\it i.e.}
$\big(-\Box_g + \frac16 R(x)\big) G(x) = \delta^{(4)}(x)$
(the symbol $\star$ stands for the convolution integral). 
The above term represents a physical effect because, 
unlike (\ref{Pol2}), it cannot be rendered local by  an appropriate
gauge choice. Unlike for the conventional treatment of Liouville theory,
we now have to deal squarely with the Lorentzian signature, as there is no equivalence 
theorem relating pseudo-Riemannian and Riemannian geometry.

To arrive at a quantitative estimate we consider a cosmological solution of the 
classical Einstein equations with metric $ds^2 = a^2(\eta) \big( -d\eta^2 + d\bx^2\big)$,
where $\eta$ is conformal time, and with the retarded Green's function \cite{Waylen}
\beq
G\big(\eta,\bx;\eta',\by \big) = \frac{1}{4\pi |\bx - \by|} \cdot 
   \frac{\delta(\eta - \eta' - |\bx - \by|)}{a(\eta) a(\eta')}
\eeq
valid for any profile of the scale factor $a(\eta)$. 
As an example let us consider the radiation era ending at $t\!=\!\trad$ and starting at
 $t_0=n_* l_P$ where $n_*$ can be large (say, $10^8$, {\it i.e.} long after the Planck era). Then
$\eta\!=\!2\sqrt{t\,\trad}$ and $a(\eta)\!=\!\eta/(2 \trad)$; furthermore $R=0$ for radiation, so the last convolution in (\ref{DesSch}) receives contributions only from 
primordial perturbations, therefore we can very crudely estimate  this contribution as 
$\na_0 \left( G \star R \right)\approx 10^{-5} \trad^{-1}$ \cite{Slava}.
However, the first convolution integral gives a dramatically bigger contribution since  in the radiation era $\cK = 3/(2 t^4)$. With $\bx=0$ and $t=\trad$, and neglecting subleading contributions we get
\beq
\!\! \frac{\pa}{\pa t}
\int\!\rd^3 y \! \int^{\eta_{\rm rad}}_{\eta_0}\!\!\rd\eta'\, \frac{a(\eta')^4}{4\pi |{\bf y}|}
\frac{\de(\eta-\eta'-|{\bf y}|)}{a(\eta)a(\eta')}\frac{3}{2 t'^4}
=\frac{8}{(n_* \lPL)^3}
\eeq
We therefore see that the pre-factor $\lPL^{-2}$ in (\ref{MEFT}) that would normally suppress
the corrections on the r.h.s. of (\ref{MEFT}) can be `beaten' by the anomalous action because the convolution integral picks up the contributions along the past lightcone back to $t_0 = n_* \lPL$, 
yielding an anomalous $T_{00}\sim 10^{-5}\trad^{-1} (n_* \lPL)^{-3}$ that with our  
assumptions on $n_*$ is much bigger than the l.h.s. $\sim \trad^{-2} \lPL^{-2}$.
Note that we do not even need to go back all the way to the singularity $t \!\sim \!  \lPL$
for this argument! In this way a perfectly smooth solution to Einstein's equation could
receive a very large `jolt' from the non-local contributions; similar `jolts' could
be expected for wave-like solutions.

Let us emphasise that no such effects are expected to arise from the non-anomalous
part of the non-local effective action which is obtained by summing 1PI 
Feynman diagrams in the usual way, or equivalently by `integrating out' massive modes.
At low energies their contribution would reduce to vertices which are effectively local
(similar to the 4-Fermi interaction in electroweak theory), and hence would remain
suppressed. By contrast, the conformal anomaly is independent of the mass of the state
that is integrated out, hence can be viewed as an IR manifestation of physics
in the deep UV region. This feature is reflected in the anomalous part of the effective 
action, whose non-localities are independent of the masses of the particles 
contributing to the anomaly, as a consequence of which the resulting corrections 
to Einstein's equations need not remain suppressed (similar non-localities would 
be present in the Standard Model if the gauge anomalies did not cancel).

Extending this argument to the full theory requires knowledge of the 
complete expression for $\Feff$ which is not available, and whose construction 
remains a formidable theoretical challenge.  Although we expect that the higher 
order terms in $\Feff$ will not affect our main conclusions, let us mention 
two possible caveats. One is that the non-localities in $\Feff$ and (\ref{MEFT}) 
could cancel between different contributions; this appears unlikely in view of the 
fact that the cancellation would have to involve infinitely many terms. 
The other caveat is that a resummation in the series expansion of $\Feff$ might 
soften or even remove the singularity of the inverse differential operators 
(conversely, it might also introduce {\em new} singularities, 
further strengthening the arguments presented above). 

\vspace{1mm}
\noindent{\bf{4. Cancelling the conformal anomaly.}} 
We now survey the theories
for which the contribution from the different spins to  the coefficients $c_s$ and 
$a_s$ cancel.
Below we tabulate the coefficients  $c_s$ and $a_s$ coming from the integration 
over massless and massive fields (first and last two columns, respectively), with spins 
up to $s\!=\!2$. These results can be excerpted from the literature, although they
are often not presented in precisely this form; for a derivation for the massless 
fields, see refs. \cite{Duff1,CD,EH,BD,FT,Vass,T}. 
%(see also \cite{LL} for a more recent application in an AdS/CFT context). 
The entries labeled $0^*$ give the results
for two-form fields (which have no massive analog); these give the same contribution 
to $c_0$ as the scalars, but their contribution to the $a_0$ coefficient is different.
The numbers for the massive fields are obtained in the usual  way by adding the lower 
helicities -- implying that spontaneous symmetry breaking does not  affect the 
conformal anomaly. 

We first  consider the {\em sum} of the anomaly coefficients.  As shown long ago  \cite{NT} 
the sum $\sum (c_s + a_s)$ can be made to vanish for all $N = 3$ supermultiplets with
maximum spin $s\leqslant 2$ by 
replacing one scalar by one two-form field in the spin-$\frac32$ multiplet; the resulting
multiplets can then be used as building blocks to arrange for all $N\geqslant 4$ supergravities 
to have vanishing $\sum (c_s + a_s)$. This implies in particular that the combined
contribution vanishes for the maximal $N=4$ Yang-Mills multiplet. 

\renewcommand{\arraystretch}{1.5}
\begin{center}
\begin{tabular}{|c||c|c||c|c|}
\hline
\multirow{2}{*} 
&\multicolumn{2}{c||}{massless}&\multicolumn{2}{c|}{massive}\\ \cline{2-5}
& $c_s$ & $a_s$& $c_s$ & $a_s$\\ \hline \hline
\  $0$($0^*$) &\ $\frac32$$(\frac32)\ $ &\ $-\frac12$($\frac{179}2$) \ &\ $\frac32$($\varnothing $)\ & \ $-\frac12$($\varnothing $)  \\[3pt] \hline
$\frac12$ & $\frac92$  & $-\frac{11}{4}$  & $\frac92$    & $-\frac{11}{4} $ \\[3pt] \hline
$1$ & $18$  & $-31$   & $\frac{39}{2} $   & $-\frac{63}{2} $ \\[3pt] \hline
$\frac32$ & $-\frac{411}{2}$   & $\frac{589}{4}$    & $-201$    & $\frac{289}{2} $ \\[3pt] \hline
$2$ & $783$   & $-571$    & $\frac{1605}{2} $   & $-\frac{1205}{2}$  \\[3pt] \hline
\end{tabular}
\vspace{2mm}

Table 1. Coefficients of the conformal anomaly in (\ref{anom})
\end{center}

Next, turning to the $C^2$ anomaly, it is a remarkable fact that the presence of gravitinos, 
hence supergravity, is mandatory for any cancellation involving the coefficients $c_s$ 
because the gravitinos are the only fields that contribute with a negative sign. 
It is easy to check that no $N\leqslant 4$ supergravity can achieve this because the 
summed contribution from the pure supergravity multiplets for $N\leqslant 4$ is 
positive, hence of the same 
sign as the one from any of the associated matter multiplets, so no cancellation of the 
$C^2$ anomaly is possible for these theories at all. One then checks that the 
total contribution $\sum c_s$ vanishes for the $N \geqslant 5$ supergravities,
\bea
c_2 + 5 c_{\frac32} + 10 c_1 + 11 c_{\frac12} + 10 c_0 &=& 0\;\nn\\
c_2 + 6 c_{\frac32} + 16 c_1 + 26 c_{\frac12} + 30 c_0 &=& 0 \; ,\nn\\
c_2 + 8 c_{\frac32} + 28 c_1 + 56 c_{\frac12} + 70 c_0 &=& 0 \;,
\eea
but for no other theories! 
Unlike the cancellation of gauge anomalies which can in principle be arranged 
for arbitrary chiral gauge groups by suitable choice of matter multiplets, the above 
cancellation thus singles out three very special theories. 

Using the results from \cite{GN} one can furthermore show that, in addition to the 
cancellation for the massless supergravity multiplets,  the contributions 
to $c_s$ also cancel for the whole tower of Kaluza-Klein states of maximal 
$D=11$ supergravity  compactified on $S^7$, and they do so `floor by floor' -- 
exactly as for the one-loop $\beta$-function in \cite{GN,Inami}!  For the reader's convenience 
we reproduce below from \cite{GN} the full table of massive Kaluza-Klein multiplets with
the Dynkin labels of the associated SO(8) representations (note that some towers start 
`higher up', because no entry in a Dynkin label can be negative).

\renewcommand{\arraystretch}{1.5}
\begin{center}
\begin{tabular}{|c|c|}
\hline
&$SO(8)$ representations\\ \hline \hline
\multirow{2}{*}{\ $0$\ } &\ $[n\!+\!2\ 0\ 0\ 0]$\,,\, $[n\ 0\ 2\ 0]$\,,\, $[n\!-\!2\ 2\ 0\ 0]$,\   \\ 
&$[n\!-\!2\ 0\ 0\ 2]$\,, $[n\!-\!2\ 0\ 0\ 0]$   \\  \hline
\multirow{2}{*}{$\frac12$} & $[n\!+\!1\ 0\ 1\ 0]$\,, $n\!-\!1\ 1\ 1\ 0]$,  \\ 
&$[n\!-\!2\ 1\ 0\ 1]$\,, $[n\!-\!2\ 0\ 0\ 1]$   \\  \hline
$1$ &\ $[n\ 1\ 0\ 0]$\,, $[n\!-\!1\ 0\ 1\ 1]$\, , $[n\!-\!2\ 1\ 0\ 0]$\    \\ \hline
$\frac32$ & $[n\ 0\ 0\ 1]$\, , $[n\!-\!1\ 0\ 1\ 0]$  \\ \hline
$2$ & $[n\ 0\ 0\ 0]$   \\ \hline
\end{tabular}

\vspace{2mm}
Table 2. Massive states in the $S^7$ compactification \cite{GN}
\end{center}

Using the $c_s$ coefficients for the massive representations from Table 1 and 
the dimension formula from \cite{GN} it is straightforward 
to prove that the sum over spins vanishes separately for each $n$. Furthermore, 
since the anomaly is independent of the chosen background solution it follows
immediately that this result remains true {\em for any compactification of M theory to four 
dimensions}. However, for other compactifications, the cancellations in general are not 
expected to work `floor by floor', but rather involve different `floors' of the Kaluza Klein tower.

From the table we see that the contribution from the spin-0 fields to the 
Gauss-Bonnet invariant $\GB$ depends on the field representation, unlike $c_0$. 
From the results of \cite{NT} it follows directly that one can arrange for the 
sum $\sum a_s$ to vanish for all $N\geqslant 5$ supergravities with appropriate
choice of field representation for the spin-0 degrees of freedom, but for no other
theories. To be sure, the contributions to $a_s$ fail to cancel 
if one chooses to represent all spin-0 degrees of freedom by 
scalar fields, and they also do not cancel for the massive Kaluza-Klein supermultiplets. 
Nevertheless, as shown in \cite{GN}, even in that case the total contribution to $a_s$ 
does cancel provided one adds up the contribution from {\em all}  Kaluza-Klein supermultiplets. 
The argument involves a $\zeta$-function regularisation and is basically due to the fact 
that one compactifies on an odd-dimensional manifold. Again, this result is background 
independent and thus remains valid for any compactification of M theory to four dimensions.

Finally, it has not escaped our attention that the complete cancellation of conformal
anomalies takes place for the very same theories for which the composite R symmetry anomaly 
cancels \cite{Marcus}.  As argued in \cite{Kallosh,Zvi} it is precisely the presence or absence 
of this anomaly that is responsible for the finiteness properties of the $N\geqslant 5$  
supergravities, and that explains the appearance of non-renormalizable divergences 
in $N=4$ supergravity at four loops. This remarkable coincidence leads us to conjecture 
that the conformal anomaly may play a similar role for the finiteness properties of extended supergravities.  Indeed, it is easy to see  that the insertion of $\Feff$ will induce new contributions
from four loops onward.

\noindent{\bf {Acknowledgments:}}  K.A.M. was supported by the Polish NCN grant 
DEC-2013/11/B/ST2/04046; he also thanks the AEI for hospitality and support
during this work. H.N. would like to thank S. Kuzenko and P. Bouwknegt 
for hospitality at UWA in Perth and ANU in Canberra, respectively, while this work 
was underway, and A. Buonanno, R. Kallosh and A. Schwimmer for discussions. 
K.A.M. would like to thank R. Penrose for discussions in Czerwi{\'n}sk and Oxford.


\begin{thebibliography}{99}
\bibitem{Deser2} S.~Deser, M.J.~Duff and C.~Isham, Nucl. Phys. {\bf B111}(1976) 45.
\bibitem{Duff1} M.J.~Duff, Nucl. Phys. {\bf B125}(1977) 334
\bibitem{CD} S.M. Christensen and M.J. Duff, Phys. Lett. {\bf 76B} (1978) 571.
\bibitem{FT} E. Fradkin and A. Tseytlin,  Nucl.Phys. {\bf B203}(1982) 157
\bibitem{DS}S.~Deser and A.~Schwimmer, Phys. Lett. {\bf B309} (1993) 279
\bibitem{Deser1} S.~Deser, Phys. Lett. {\bf B 479} (2000) 315
%Helv.Phys.Acta {\bf 69} (1996) 570
\bibitem{T} A.~Tseytlin, Nucl. Phys. {\bf B877} (2013) 598
\bibitem{Mott} P.O.~Mazur and E.~Mottola, Phys.Rev. {\bf D64}(2001)104022
\bibitem{Mott1} E.~Mottola, {\tt arXiv:1606.09220} 
\bibitem{P} A.M.~Polyakov, Phys. Lett. {\bf B103} (1981) 207
\bibitem{DK} J.~Distler and H.~Kawai,
%Conformal Field Theory and 2D Quantum Gravity 
                      Nucl.Phys. {\bf B321} (1989) 509     %-527 
\bibitem{Davies} P.C.W. Davies, Phys. Lett. {\bf B68} (1977) 402
\bibitem{Hartle} M. V. Fischetti, J.B. Hartle and B.L.Hu, Phys. Rev. {\bf D20} (1979) 1757                      
\bibitem{Tr} A. Trautman A, Bull. Acad. Polon. Sci. {\bf 6} (1958) 407 
%(reprinted as arXiv:1604.03145)
\bibitem{Waylen} P.C.~Waylen, Proc.R.Soc.London, {\bf A362} (1978) 233                      
\bibitem{Slava} See {\it e.g.} V.~Mukhanov, {\it Physical Foundations of Cosmology},  
              Cambridge University Press, 2005. The fact that $R$ vanishes in the radiation
              era is not an important feature here, as there are other terms in $\Feff$
              that depend only on the Ricci and Riemann tensors (which do not vanish).
%\bibitem{KM1} K.A.~Meissner and H.~Nicolai, Phys. Lett. {\bf B660} (2008) 260
\bibitem{EH}  T. Eguchi, P. B. Gilkey, and A. J. Hanson, 
%Gravitation, Gauge Theories and Differential Geometry, 
       Phys. Rept. {\bf 66} (1980) 213.
\bibitem{BD}  N. D. Birrell and P. C. W. Davies, 
    {\it Quantum Fields in Curved Space}, Cambridge University Press, 1982.
\bibitem{Vass} D. Vassilevich, %Heat kernel expansion: User's manual, 
            Phys.Rept. {\bf 388} (2003) 279%??360   %, [hep-th/0306138].
%\bibitem{LL} F. Larsen and P.~Lisbao, JHEP {\bf 1601} (2016) 024
\bibitem{NT}H.~Nicolai and P.K.~Townsend,
             Phys.Lett. {\bf B98} (1981) 257-260 
\bibitem{GN} G.W. Gibbons and H. Nicolai, Phys. Lett. {\bf 143B} (1984) 108.             
\bibitem{Inami} T.~Inami and K.~Yamagishi, Phys. Lett. {\bf B143} (1984) 115                         
\bibitem{Marcus} N.~Marcus, Phys.Lett. {\bf B157} (1985) 383                                    
\bibitem{Kallosh} J.J.M. Carrasco, R. Kallosh, R. Roiban and A.A. Tseytlin, 
%On the U(1) duality anomaly and the S-matrix of N=4 supergravity 
          JHEP {\bf 1307} (2013) 029 
\bibitem{Zvi} Z. Bern, S. Davies and T. Dennen,
%Enhanced ultraviolet cancellations in N=5 supergravity at four loops 
                  Phys.Rev. {\bf D90} (2014) 105011          
%\bibitem{CDGR} S.M.~Christensen, M.J.~Duff, G.W.~Gibbons and M.~Rocek,
 %      Phys. Rev. Lett. {\bf 45} (1980) 161

\end{thebibliography}
\end{document}